\input{aipcheck}

\documentclass[
   ,final            
  ]
  {aipproc}

\layoutstyle{8x11double}
\renewcommand\XFMtitleblock{%
  \XFMtitle
  \let\XFMoldpar\par
  \def\par{\XFMoldpar\def\par{\space 
             (on behalf of the MAGIC Collaboration)\XFMoldpar}}%
   \XFMauthors
   \let\par\XFMoldpar
   \XFMaddresses
   \XFMabstract
   \vspace{5pt}%
   \XFMkeywords
   \XFMclassification
 }

\usepackage{epstopdf}
\usepackage{graphicx} 

\def\ea{\emph{et al.}~}
\def\ApJ{Astrophys. J.}
\def\AA{Astron. Astrophys.}
\def\MNRAS{Mon. Not. R. Astron. Soc.}

\begin{document}

\title{Discovery of very high energy gamma-rays from the flat spectrum radio quasar 3C~279 with the MAGIC telescope}


\author{M. Errando}{
  address={IFAE, Edifici Cn., Campus UAB, E-08193 Bellaterra, Spain}
}
\author{R. Bock}{
  address={Max-Planck-Institut f\"ur Physik, D-80805 M\"unchen, Germany},altaddress={Universit\`a di Padova and INFN, I-35131 Padova, Italy}
}
\author{D. Kranich}{
  address={ETH Zurich, CH-8093 Switzerland}
}
\author{E. Lorenz}{
  address={ETH Zurich, CH-8093 Switzerland},altaddress={Max-Planck-Institut f\"ur Physik, D-80805 M\"unchen, Germany}
}
\author{P. Majumdar}{
  address={DESY Deutsches Elektr.-Synchrotron D-15738 Zeuthen, Germany}
}
\author{M. Mariotti}{
  address={Universit\`a di Padova and INFN, I-35131 Padova, Italy}
}
\author{D. Mazin}{
  address={IFAE, Edifici Cn., Campus UAB, E-08193 Bellaterra, Spain}
}
\author{E. Prandini}{
  address={Universit\`a di Padova and INFN, I-35131 Padova, Italy}
}
\author{F. Tavecchio}{
  address={INAF National Institute for Astrophysics, I-00136 Rome, Italy}
}
\author{M. Teshima}{
  address={Max-Planck-Institut f\"ur Physik, D-80805 M\"unchen, Germany}
}
\author{R. Wagner}{
  address={Max-Planck-Institut f\"ur Physik, D-80805 M\"unchen, Germany}
}

\begin{abstract}
3C~279 is one of the best studied flat spectrum radio quasars located
at a comparatively large redshift of $z=0.536$. Observations in the very
high energy band of such distant sources were impossible until
recently due to the expected steep energy spectrum and the strong
gamma-ray attenuation by the extragalactic background light 
photon field, which conspire to make the source visible only with a
low energy threshold. Here the detection of a significant
gamma-ray signal from 3C~279 at very high energies ($E>75$\,GeV) during a flare in early
2006 is reported. Implications of its energy spectrum on the current understanding
of the extragalactic background light and very high energy gamma-ray emission mechanism models are
discussed.
\end{abstract}

\maketitle


\section{Introduction}

The flat spectrum radio quasar (FSRQ) 3C~279, located at a redshift of $z=0.536$, is one of the brightest extragalactic objects in the gamma-ray sky \cite{kniffen93} 
and it is an exceptionally variable
source at various energy bands, including very strong gamma-ray flares recorded by satellite detector EGRET in 1991 and 1996 \cite{Wehrle98}. Its detection at very high energy (VHE, defined here as $E>50$\,GeV) gamma-rays by the MAGIC telescope \cite{icrc,science} is the first detection of an FSRQ in this energy band as well as the farthest object detected at these wavelengths, as the highest redshift previously observed was $z=0.212$ \cite{magic07}. 

The EGRET experiment 
detected 66 blazars: 51 FSRQ and 15
BL Lac objects \cite{hartman}. Blazars are thought to be powered by accretion of matter onto supermassive black holes residing in the centers of galaxies and ejecting relativistic jets with small angles to the line of sight \cite{Blandford72}. 
Except for the FSRQ 3C~279, all the blazars detected at VHE up to know are of the BL Lac type, being the difference between this two subclasses that FSRQs show optical emission lines while in BL Lacs these are very weak or absent.
The spectral energy distribution (SED) of blazars is characterized by two broad bumps, the first one peaking between the infrared and the X-ray band and the second one in the gamma-ray domain. The first peak is usually attributed to synchrotron emission from relativistic electrons, while the high-energy peak is presumably due to the inverse Compton scattering of low-energy photons by these electrons. The soft photon population can be the synchrotron photons themselves (synchrotron self-Compton or SSC mechanism \cite{Maraschi92}) or ambient radiation (external Compton or EC mechanism \cite{dermer93,sikora94}). Other scenarios (hadronic models) involve gamma-ray emission due to accelerated protons and ions \cite{mannheim92,mucke03}.

Apart from being the first FSRQ detected in the VHE band, the detection of 3C~279 is important because gamma-rays from distant sources are expected to be strongly attenuated in intergalactic space by the possible interaction with low-energy photons ($\gamma + \gamma \rightarrow e^+ + e^-$). These photons form the extragalactic background light (EBL)  \cite{Hauser01} and have been radiated by stars and galaxies in the course of cosmic history. The attenuation by the EBL increases with increasing distance to the source and energy of the emitted gamma-ray, therefore, VHE radiation from distant sources is strongly suppressed at high energies (for $E\geq 300$\,GeV and $z=0.536$ more than 90\% of the emitted gamma-rays are expected to be absorbed).

\section{Observations and Analysis}

Observations of 3C~279 were performed using the MAGIC telescope (see \cite{crab} for a detailed
description), which is located on the Canary Island of La Palma and has a 17\,m-diameter tessellated reflector dish and a photomultiplier camera covering $3.5^\circ$ field of view. 
It has a trigger threshold of 60\,GeV, an angular
resolution of approximately 0.1$^\circ$ and an energy resolution
above 150\,GeV of about 25\%.

Data were taken in \textit{on/off mode}, which includes \textit{on} events taken with the telescope pointing at the source and a similar amount of  \textit{off} events taken at a sky location near the source with very similar operating conditions.
3C~279 was observed over 10 nights between January and April 2006 for a total of $9.7$ hours. Simultaneous optical R-band observations with the 1.03-m telescope at the Tuorla Observatory (Finland) and the 35-cm KVA telescope on La Palma revealed that during MAGIC observations 3C~279 was in high optical state, a factor of 2 above its long-term baseline flux (host galaxy subtracted).

The data anaysis was carried out using the standard MAGIC analysis and reconstruction software \cite{crab}. To search for a gamma-ray signal, first the standard calibration of the data was
performed. Then an image cleaning procedure was applied
using the amplitude of the calibrated
signals. The next step included
the parameterization of the shower images \citep{Hillas1985}.
Hadronic background suppression was achieved
using the Random Forest (RF) method \citep{Breiman2001, Bock2007},
where for each event the so-called hadronness (\textit{h}) parameter is
computed based on the image parameters. 
Moreover, the RF method
was also used for the energy estimation using a Monte Carlo simulated
gamma-ray sample with the same zenith angle distribution as the data
sample.
All image parameters were checked for consistency between \textit{on} and \textit{off} data. 

\section{Results}
The excess events were obtained by subtracting suitably normalized \textit{off} from the \textit{on} data. Cuts in \textit{h} and \textit{alpha} (the latter describes the direction of the main axis of the image) were optimized using Crab Nebula data taken at similar zenith angles. For the detection and light curve the cuts $h<0.12 $ and $alpha<12^{\circ}$ where used, resulting in a combined efficiency of 40\% for gammas. For these cuts and a quality cut that removed small images the resulting energy threshold  is 170~GeV.

\begin{figure}
  \includegraphics[width=0.75\columnwidth]{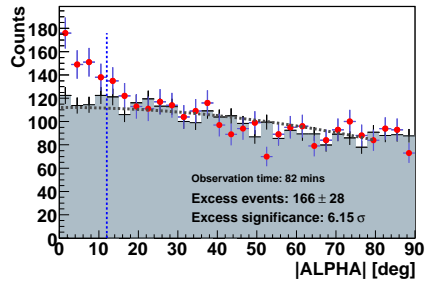}
  \caption{Histogram of the image parameter \textit{alpha} obtained for 3C~279 in the night of 23 February 2006 for \textit{on} (dots) and \textit{off} (crosses and filled area) pointings. The dotted line is a simple parabolic fit without linear term to the \textit{off} events between 0$^{\circ}$ and 80$^{\circ}$, serving to smoothen the background distribution for a better extrapolation towards $alpha=0$. The number of excess events is obtained with respect to this line.}
  \label{alpha}
\end{figure}

\begin{figure}
  \includegraphics[width=0.75\columnwidth]{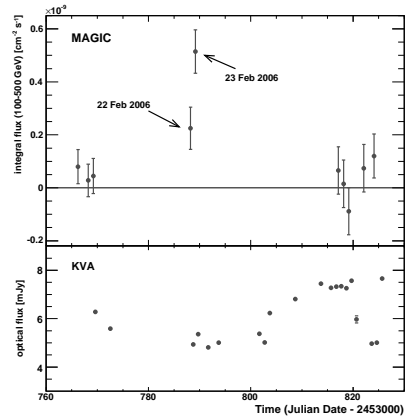}
  \caption{MAGIC (top) and optical R-band (bottom) light curves obtained for 3C~279 from February to March 2006. The long-term baseline for the optical flux is 3\,mJy (host galaxy subtracted).}
  \label{lc}
\end{figure}

3C~279 was clearly detected on the night of 23 February 2006 (Fig.~\ref{alpha}) with a significance of 6.2$\sigma$, which translates into 5.8$\sigma$ when corrected with a trial factor corresponding to 10 nights of observation. A less significant excess on the previous night is visible in the light curve (Fig.~\ref{lc}). As determined by the $\chi^{2}$ test, the probalility that the gamma-ray flux on all 10 nights was zero is $2.3 \times 10^{-7}$, corresponding to 5.0$\sigma$ in a Gaussian distribution.

\begin{figure}
  \includegraphics[width=0.75\columnwidth]{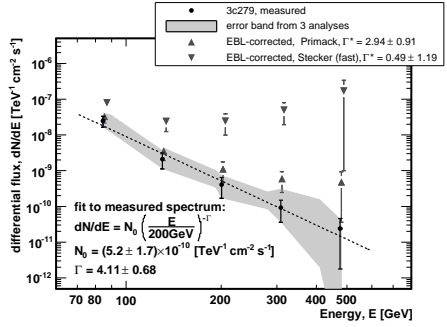}
  \caption{Differential energy spectrum of 3C~279 measured by MAGIC. The grey area includes the combined statistical (1$\sigma$) and systematic errors, and underlines the marginal significance of detections at high energy. The dotted line shows compatibility of the measured spectrum with a power law of photon index $\alpha = 4.1$. The triangles are measurements corrected on the basis of the two models for EBL density discussed in the text.}
  \label{spectrum}
\end{figure}

The observed differential VHE spectrum (Fig.~\ref{spectrum}) was obtained with looser cuts to increase the number of gamma-ray candidates, and can be described by a power law with spectral index of $\alpha = 4.1 \pm 0.7_{\mathrm{stat}} \pm 0.2_{\mathrm{syst}}$. The measured integrated flux above 100~GeV on 23 February is $\left(5.15 \pm 0.82_{\mathrm{stat}} \pm 1.50_{\mathrm{syst}} \right) \times 10^{-10}$\,cm$^{-2}$\,s$^{-1}$.

\section{Discussion}

\begin{figure}
  \includegraphics[width=0.95\columnwidth]{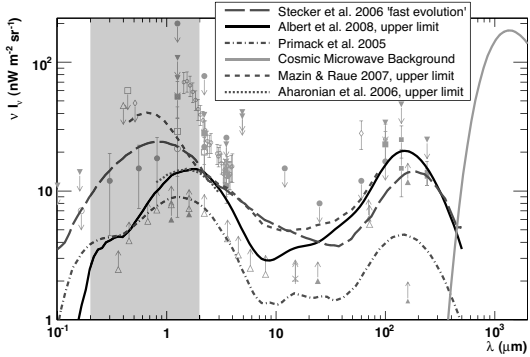}
  \caption{EBL models at $z=0$ and direct measurements of the EBL photon density at various wavelengths. The upper limit derived from the detection of 3C~279 (black line) is based on \cite{Kneiske02} with parameters adapted to the results from the latest galaxy counts (see Supporting Material in \cite{science} for details). The dotted line is an upper limit from \cite{Mazin07}. The shaded region corresponds to the range of frequencies where the MAGIC measured spectrum is sensitive.}
\label{ebl}
\end{figure}

The EBL influences the propagation of gamma-ray emitted from distant sources resulting in an exponential decrease with energy and a cutoff in the measured gamma-ray spectrum.
Several models have been proposed for the EBL \cite{Hauser01}. All have limited predictive power for the EBL density, particularly as a function of the redshift, as many details about star and galaxy evolution remain uncertain. To illustrate this uncertainty two extreme models are used to calculate the source intrinsic spectrum: a model by Primack \ea \cite{Primack05} close to the lower limits set by galaxy counts \cite{Madau00,Fazio04} and the `fast-evolution' model by Stecker \ea \cite{Stecker06} corresponding to the highest possible attenuation. These models will be referred as \textit{low} and \textit{high} respectively and are shown in Figure~\ref{ebl}.

\begin{figure}
  \includegraphics[width=0.85\columnwidth]{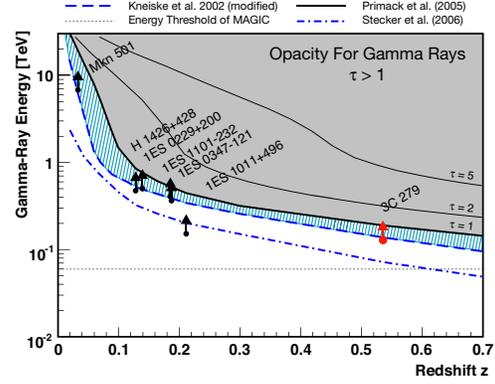}
  \caption{The gamma-ray horizon. The redshift region over which the gamma-ray horizon can be constrained by observations has been extended up to z = 0.536. The prediction range of EBL models is illustrated by \cite{Primack05} (thick solid line) and \cite{Stecker06} (dashed-dotted line). The tuned model of \cite{Kneiske02} (dashed line) represents an upper EBL limit based on our 3C 279 data, obtained on the assumption that the intrinsic photon index is $\geq 1.5$ (arrow corresponding to 3C~279). Limits obtained for other sources are shown by black arrows, most of which lie very close to the model \cite{Kneiske02}. The narrow filled band is the region allowed between this model and a maximum possible transparency (i.e., minimum EBL level) given by \cite{Primack05}, which is nearly coincident with galaxy counts. The gray area indicates an optical depth $\tau > 1$, where the flux of gamma rays is strongly suppressed. To illustrate the strength of the attenuation in this area energies for $\tau = 2$ and $\tau = 5$ (thin black lines) are shown, again with \cite{Primack05} as model.}
  \label{horizon}
\end{figure}

A power-law fit to the EBL-corrected spectral points results in an intrinsic photon index $\alpha^{\star}=2.9 \pm 0.9_{\mathrm{stat}} \pm 0.5_{\mathrm{syst}}$ (\textit{low} absorption) and $\alpha^{\star}=0.5 \pm 1.2_{\mathrm{stat}} \pm 0.5_{\mathrm{syst}}$ (\textit{high} absorption).
The \textit{high} model leads to an intrinsic VHE spectrum difficult to reconcile with an extrapolation of the EGRET data and with general constraints in the spectral energy distribution. On the other hand, the \textit{low} model apparently gives an acceptable result.
The distance at which the flux of photons of a given energy is attenuated by a factor $e$ (the path corresponding to an optical depth $\tau=1$) is called the gamma-ray horizon \cite{Fazio70} and is usually expressed as a function of the redshift. This energy-redshift relation is shown in Figure~\ref{horizon}. 

Assuming a hardest intrinsic photon index $\alpha^{\star}=1.5$ a model based on Kneiske \ea \cite{Kneiske02,Kneiske04} can be tuned to give an upper limit to the EBL density in the $0.2 - 2$\,$\mu$m range as shown in Fig.~\ref{ebl}. This result supports the conclusion drawn from VHE measurements at lower redshifts \cite{Aharonian06} that the EBL density is close to the lower limits set by galaxy counts. It is important to point out that this limit is only valid under the assumption that the intrinsic photon index cannot be harder than 1.5, which is the hardest value given for EGRET sources and is also the hardest that can be obtained with classical leptonic emission mechanisms. Alternative scenarios can produce intrinsic spectra with $\alpha^{\star}<1.5$. SSC models with a narrow electron distribution can produce spectra with $\alpha^{\star} \approx 0.7$ \cite{kata06}, and internal absorption by soft photons can also harden the spectrum \cite{bednarek97,aharonian08,liu08,bednarek08}. However, an accurate modeling of the internal absorption in 3C~279 finds no important hardening of its spectrum in the energy band measured by MAGIC \cite{mazin08}.

\begin{figure}
  \includegraphics[width=0.75\columnwidth]{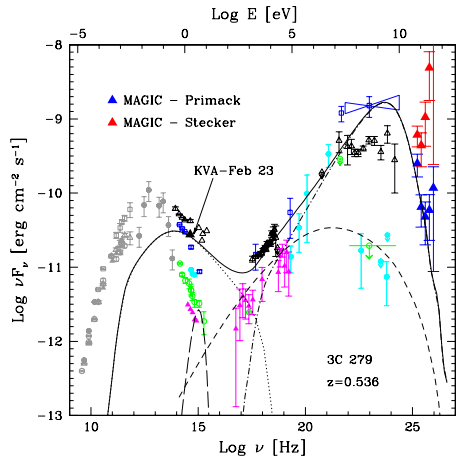}
  \caption{Spectral energy distribution for 3C~279 at different source intensities, over the years from 1991 to 2003, taken from \cite{Ballo02,Collmar04} with MAGIC points (2006) at the high-energy end. The MAGIC points are corrected according to the two extreme EBL models discussed on the text. The spectrum corrected with the \textit{low} EBL model gives clearly a more plausible result. An emission model and its individual components are also shown: the solid line represents the total model emission, fitted only to the MAGIC points unfolded with a \textit{low} EBL density and the simultaneous optical point from the KVA telescope. The individual components of the emission model are synchrotron radiation (dotted line), disk emission (long-dashed), synchrotron-self-Compton (short-dashed) and external Compton (dot-dashed).}
  \label{sed}
\end{figure}

The emission mechanism responsible for the observed VHE radiation remains uncertain. The spectrum corrected with a \textit{low} level EBL can be reproduced by a one-zone SSC+EC model \cite{maraschi03} (Fig.~\ref{sed}) while the spectrum corrected with a \textit{high} absorption can not be fitted by generic leptonic emission models. Studies including quasi-simultaneous optical and X-ray data \cite{webt} conclude that neither one-zone SSC nor one-zone EC models can reproduce the observed SED \cite{bottcher}.
 
\begin{theacknowledgments}
MAGIC enjoys the excellent working conditions at the ORM in La Palma and is supported by the German BMBF and MPG, the Italian INFN, the Spanish MCI-NN, ETH research grant TH 34/04 3, the Polish MNiSzW Grant N N203 390834 and by the YIP of the Helmholz Gemeinschaft.
\end{theacknowledgments}

\bibliographystyle{aipproc}   

\end{document}